\documentclass[structabstract]{aa}  
\usepackage{graphicx}
\usepackage{txfonts}
\begin{document}
\title{The nearby eclipsing stellar system $\delta$ Velorum}
\subtitle{III. Self-consistent fundamental parameters and
  distance\thanks{Based on observations made with ESO telescopes at
    Paranal Observatory, under ESO programs 076.D-0782(B),
    081.D-0109(B), 081.D-0109(C), 282.D-5006(A) and Arcetri GTO
    program 084.C-0170(C)}}

\author{
  A.~M\'erand \inst{1}
  \and P.~Kervella \inst{2}
  \and T.~Pribulla\inst{3,4}
  \and M.~G.~Petr-Gotzens\inst{5}
  \and M.~Benisty\inst{6} 
  \and A.~Natta\inst{7,8} \and 
  G. Duvert \inst{9} \and
  D. Schertl\inst{10} \and
  M. Vannier \inst{11} }

\institute{European Southern Observatory, Alonso de Cordova 3107,
  Casilla 19001, Vitacura, Santiago 19, Chile \and LESIA, Observatoire
  de Paris, CNRS\,UMR\,8109, UPMC, Universit\'e Paris Diderot, 5 place
  Jules Janssen, 92195 Meudon, France \and Astronomical Institute
  Slovak Academy of Sciences 059 60 Tatranska Lomnica, Slovak Republic
  \and Astrophysikalisches Institut und Universit\"ats-Sternwarte,
  Schillerg\"a{\ss}chen 2-3, 07745 Jena, Germany \and European
  Southern Observatory, Karl-Schwarzschild-Str. 2, D-85748 Garching,
  Germany \and Max-Planck-Institut f\"ur Astronomie, K\"onigstuhl 17,
  D-69117 Heidelberg, Germany \and Osservatorio Astrofisico di
  Arcetri, INAF, Largo E.Fermi 5, I-50125 Firenze, Italy \and School
  of Cosmic Physics, Dublin Institute for Advanced Studies, Dublin 2,
  Republic of Ireland \and UJF-Grenoble 1/CNRS-INSU, Institut de
  Plan\'etologie et d'Astrophysique de Grenoble (IPAG) UMR 5274,
  Grenoble, F-38041, France \and Max-Planck-Institut f\"ur
  Radioastronomie, Auf dem H\"ugel 69, 53121 Bonn, Germany \and
  Laboratoire Fizeau, Universit\'e de Nice, CNRS-Observatoire de la
  C\^ote d'Azur, 06108 Nice Cedex 2, France } \offprints{A. M\'erand}
\mail{amerand@eso.org}

\date{Received ---; accepted ---}

% \abstract{}{}{}{}{} 
% 5 {} token are mandatory

\abstract
%% context heading (optional) %%%%%%%%%%%%%%%%%%%%%%%%%%%%%%%%%%%
{The triple stellar system $\delta$\,Vel (composed of two
    A-type and one F-type main sequence stars) is particularly
  interesting as it contains one of the nearest and brightest
  eclipsing binaries. It therefore presents a unique
  opportunity to determine independently the physical
  properties of the three components of the system, as well as its
  distance.}
% aims heading (mandatory)%%%%%%%%%%%%%%%%%%%%%%%%%%%%%%%%%%%%%%%%
{We aim at determining the fundamental parameters (masses, radii,
  luminosities, rotational velocities) of the three components of
  $\delta$\,Vel, as well as the parallax of the system, independently
  from the existing \emph{Hipparcos} measurement.}
% methods heading (mandatory) %%%%%%%%%%%%%%%%%%%%%%%%%%%%%%%%%%%
{We determined dynamical masses from high-precision astrometry of the
  orbits of Aab-B and Aa-Ab using adaptive optics (VLT/NACO) and
  optical interferometry (VLTI/AMBER). The main component is an
  eclipsing binary composed of two early A-type stars in rapid
  rotation. We modeled the photometric and radial velocity
  measurements of the eclipsing pair Aa-Ab using a self consistent
  method based on physical parameters (mass, radius, luminosity,
  rotational velocity).}
% results heading (mandatory) %%%%%%%%%%%%%%%%%%%%%%%%%%%%%%%%%%
{From our self-consistent modeling of the primary and secondary
  components of the $\delta$\,Vel A eclipsing pair, we derive their
  fundamental parameters with a typical accuracy of 1\%. We find that
  they have similar masses, respectively $2.43 \pm 0.02$\,$M_\odot$
  and $2.27 \pm 0.02$\,$M_\odot$. The physical parameters of the
  tertiary component ($\delta$\,Vel B) are also estimated, although to
  a lower accuracy. We obtain a parallax $\pi = 39.8 \pm 0.4$\,mas for
  the system, in satisfactory agreement ($-1.2\,\sigma$) with the
  \emph{Hipparcos} value ($\pi_{\rm Hip} = 40.5 \pm 0.4$\,mas).}
{The physical parameters we derive represent a consistent set of
  constraints for the evolutionary modeling of this system. The
  agreement of the parallax we measure with the \emph{Hipparcos} value
  to a 1\% accuracy is also an interesting confirmation of the true
  accuracy of these two independent measurements.}


\maketitle
%
%________________________________________________________________

% ############################# 1 ############################
\section{Introduction}
% ############################################################

Early-type main sequence stars exhibit a number of peculiarities
usually not encountered in cooler stars: fast rotation, debris disks,
enhanced surface metallicities (Am), magnetic fields and rapid
oscillations (Ap and roAp stars), etc. Although stellar structure and
evolution models are now rather successful in reproducing the observed
physical properties of most A-type stars, the observational
constraints on these models remain relatively weak, occasionally
leading to surprising discoveries. An example is provided by the
recent interferometric observations of the A0V benchmark star
\object{Vega}, that confirmed that Vega, as previously shown by
Gulliver, Hill \& Adelman (\cite{gulliver94}), is a pole-on fast
rotator near critical velocity (Aufdenberg et
al.~\cite{aufdenberg06}). The same interferometric observations showed
that Vega harbors a hot debris disk within within 8\,AU from the star
(Absil et al.~\cite{absil06}).

\object{$\delta$ Vel} (\object{HD 74956}, \object{HIP 41913},
\object{GJ 321.3}, \object{GJ 9278}) is a bright multiple star
including at least three identified components, and is among our
closest stellar neighbors, with a revised \emph{Hipparcos} parallax of
$\pi_{\rm Hip} = 40.5 \pm 0.4$\,mas
(van~Leeuwen~\cite{vanleeuwen07}). This object has many observational
peculiarities. Firstly, it was discovered only in 1997 that
$\delta$\,Vel hosts one of the brightest of all known eclipsing
binaries (Otero et al.~\cite{otero00}), with a remarkably long orbital
period ($P \approx 45$\,days). This eclipsing binary is also one of
the very few that are easily observable with the naked eye ($m_V
\approx 2$). The eclipsing pair was first resolved using optical
interferometry by Kellerer et al.~(\cite{kellerer07}). Secondly,
$\delta$\,Vel is known to have a moderate thermal infrared excess
(e.g. Aumann~\cite{aumann85}, Su et al.~\cite{su06}), and
\emph{Spitzer} observations revealed a spectacular bow shock caused by
the motion of $\delta$\,Vel in a dense interstellar cloud (G\'asp\'ar
et al.~\cite{gaspar08}). The presence of interstellar material was
also reported by Hempel \& Schmitt~(\cite{hempel03}), who observed two
red-shifted absorbing components in absorption in the Ca~II~K line, of
probable interstellar origin. In Paper~I of the present series,
Kervella et al.~(\cite{kervella09}) confirmed that the infrared excess
is essentially emitted by the bow shock, and not warm circumstellar
material located close to the stars. In the framework of a search for
resolved emission due to debris disks, Moerchen et
al.~(\cite{moerchen10}) obtained thermal infrared images of
$\delta$\,Vel using the Gemini South telescope and the T-ReCS
instrument, and detected a marginally resolved emission at $\lambda =
10.4\,\mu$m.

In Paper~II, Pribulla et al.~(\cite{pribulla11}) used a combination of
high-resolution spectroscopy and photometric observations (from the
SMEI instrument, attached to the Coriolis satellite) to derive an
accurate orbital solution for the eclipsing binary $\delta$\,Vel~A,
and estimate the physical parameters of $\delta$\,Vel~Aa and Ab. They
identified that the two eclipsing components are fast rotating stars,
with respective masses of $2.53 \pm 0.11\,M_\odot$ and $2.37 \pm
0.10\,M_\odot$ ($\approx 4\%$ accuracy), and estimated the mass of
$\delta$\,Vel~B to be $\approx1.5\,M_\odot$.

In spite of this recent progress, uncertainties remain on the
fundamental parameters of the different components of the system, in
particular on their exact masses. Taking advantage of the availability
of NACO astrometry of the $\delta$\,Vel A-B pair, and new
interferometric observations from the VLTI/AMBER instrument, we
propose here to revisit the system along two directions. In
Sect.~\ref{AaAborbit}, we describe our new VLTI/AMBER interferometric
data, as well as our re-analysis of the spectroscopic and photometric
data previously used in Paper~II. Sect.~\ref{modelfitting} is
dedicated to the description of our self-consistent model, and the
derivation of an improved orbital solution, physical parameters as
well as an independent distance. In Sect.~\ref{ABorbit} we employ NACO
astrometry of the visual $\delta$\,Vel~A-B binary to obtain an
improved orbital solution. Compared to our work presented in Paper~II,
this new analysis result is a clearer view and better confidence in
the derived fundamental parameters of the system (for all three
components), thanks to the redundant nature of our data and our
independent determination of the distance.

%\begin{itemize} \item astrometric observations of $\delta$~Vel
%Aab-B. These observations should lead to a better determination of
%the orbit, hence a refined estimation of the total mass. Combined
%with broad-band photometry observations of the B component, we can
%infer the mass of B, hence the mass of Aab.  \item Interferometric
%data of Aab which enables us to make an orbital fit and to assess the
%mass of Aab, independently of the first step.  \item A self
%consistent modeling of photometric and spectroscopic observations of
%the eclipsing pair Aa-Ab. We use a model of the system using physical
%quantities (masses, distance, $v\sin i$, angular semi major axis,
%etc.) to constrain the physical parameters of Aa and Ab: linear
%radii, luminosities, rotational rates.  \end{itemize}

% ############################# 3 ############################
\section{The orbit and parameters of $\delta$\,Vel Aa and Ab \label{AaAborbit}}
% ############################################################ We
% gathered a broad set of of observational data on $\delta$\,Vel Aab,
% including photometry, spectroscopy (radial velocity) and
% interferometry.

\subsection{Observations and data analysis}

\subsubsection{Interferometry}

AMBER (Petrov et al.~\cite{petrov07}), the three-telescope beam
combiner of the VLTI, has the proper angular resolution to resolve the
Aa-Ab pair. This instrument combines simultaneously 3 ATs
  (Auxiliary Telescopes) or 3 UTs (Unit Telescopes) of the VLT and
  operates in the near infrared (H and K band). It has a choice of
  spectral resolutions of $R\sim35$, $R\sim1500$ or $R\sim15000$. For
  this study, we had data in low resolution (H+K bands at $R\sim35$)
  and medium resolution (H or K band, at $R\sim1500$).  We used
  baselines of the order of 100\,m in order to obtain spatial
  resolution in the milli-arcsecond regime. These interferometric
data have been collected in a dedicated program (ESO program
076.D-0782), as well as during Guaranteed Time (GTO) from Arcetri
Observatory. We present here a reduction of these data.

We reduced the data using the AMBER reduction package \texttt{amdlib3}
(Chelli et al.~\cite{chelli09}, Tatulli et al.~\cite{tatulli07}) and
performed the calibration using stellar calibrators chosen in the
catalogue by M\'erand et al.~(\cite{merand05}) and a custom software
which estimates and interpolates the transfer function of the
instrument. For each night, we derived the separation of Aa and Ab
using a $\chi^2$ map as a function of the separation vector (two
parameters). The other parameters, such as flux ratio or individual
diameters, were set using simple hypothesis and their choice did not
affect significantly our final estimated angular separations. The
resulting separation vectors are listed in
Table~\ref{tab:AMBER_positions} in coordinate towards East and North
(which correspond to the $u$ and $v$ axes in the projected baselines
map). The error bars on this vectors were estimated in the
$\chi^2$ map.

%{\bf ADD AN EXAMPLE OF $\chi^2$ MAP }

\begin{table}
  \caption{AMBER separation vectors (primary to
      secondary) \label{tab:AMBER_positions}}
  \begin{center}
    \begin{tabular}{lcc}
      \hline \hline
      date   & toward East  & toward North \\
      MJD & mas & mas\\
      \hline
       53427.09  & $-4.1\pm0.4$  & $ 7.3\pm0.5$ \\   % dataset=1; New 
       53784.08  & $-6.5\pm0.5$  & $13.8\pm0.3$ \\   % dataset=2; new
       54819.25  & $-7.3\pm0.2$  & $16.1\pm0.2$ \\   % dataset=5; new
       54832.12  &  $2.7\pm0.7$  & $-6.4\pm0.4$ \\   % dataset=4; new
       55147.38  &  $1.6\pm0.5$  & $-4.0\pm0.5$ \\   % dataset=6; new
      \hline
    \end{tabular}
    \tablefoot{For simplicity, error bars have been estimated along E
      and N directions even though the true error is an ellipse with
      orientation dicted by the $\chi^2$ map. It turns out the
      ellipses have small flattening, so the approximation is
      relevant.}
\end{center}
\end{table}

\subsubsection{Photometry}

We used the photometric data from the SMEI satellite, presented in
Paper~II. The available quantity is the relative flux normalized to
the value outside of the eclipses, since there was no absolute
calibration of SMEI data available. We corrected for the presence of
the B component which is in the field of view of SMEI. From our model
of B, the expected flux ratio between B and Aab is 7.5\% in the SMEI
bandpass. The transmission of the instrument has a triangular shape
that peaks at 700\,nm, with a quantum efficiency around 47\%, and
falls to $\approx5\%$ at 430\,nm towards the blue, and 1025\,nm
towards the red (Spreckley \& Stevens~\cite{spreckley08}). We removed
this contribution which, if not taken into account, would result in an
underestimation of the depth of the eclipses. We also incorporated to
our photometric dataset the photometric measurement we derived of Aab
in the K band (Paper~I). We use this value in our fit as the only
constraint in term of absolute photometry.

\subsubsection{Spectroscopy}

%\texttt{details}

The observables we derived from the visible spectroscopic
data are the broadening functions (BF) presented in paper II (see this
reference for more explanations). These functions contain a
lot of information: not only do they contain the radial
velocities that result from the orbital motion, but also the
broadening due to the stellar rotation and the flux ratio in the
considered band.

\begin{figure}
\begin{center}
  \includegraphics[width=0.45\textwidth]{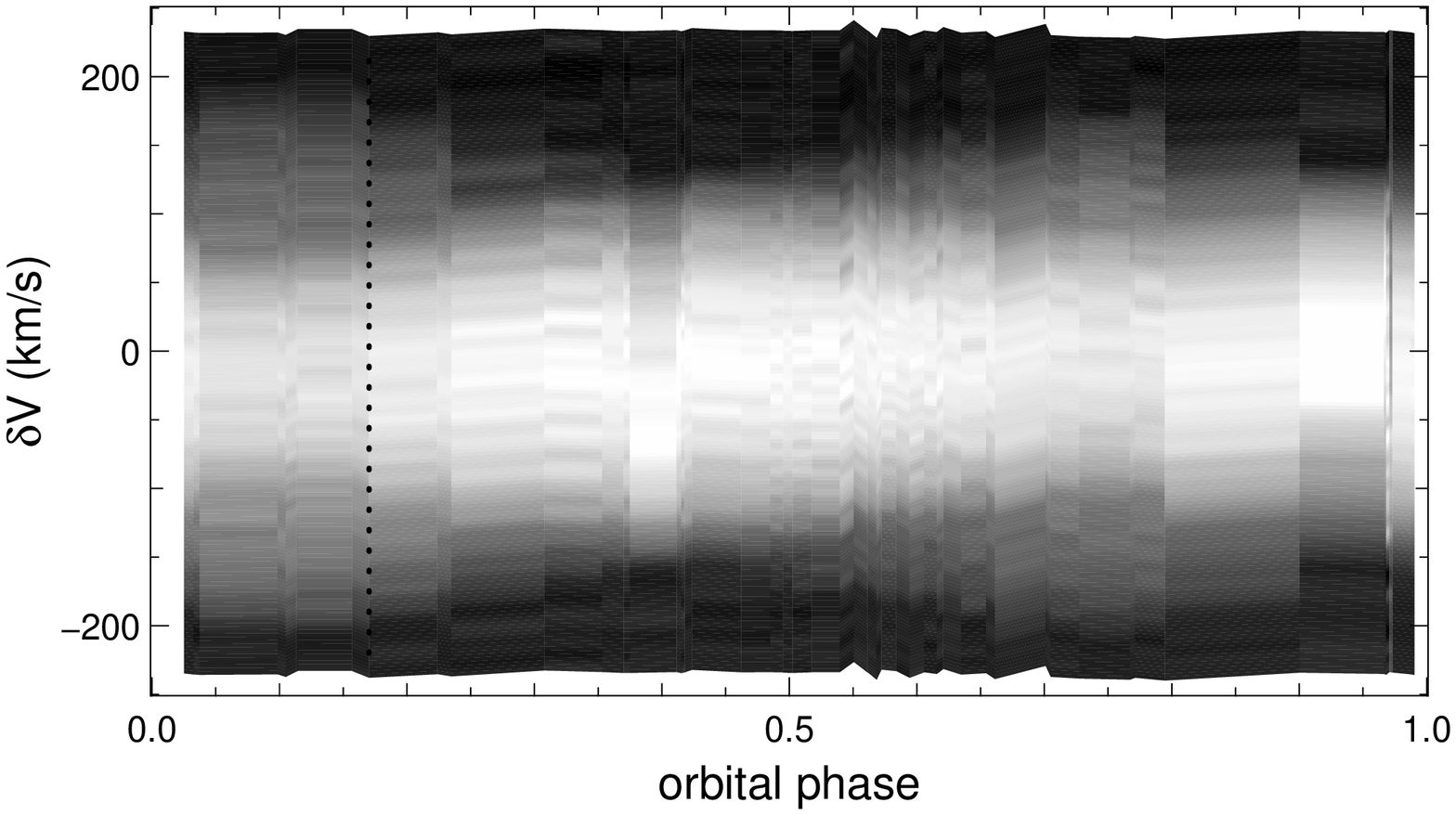}
  \includegraphics[width=0.45\textwidth]{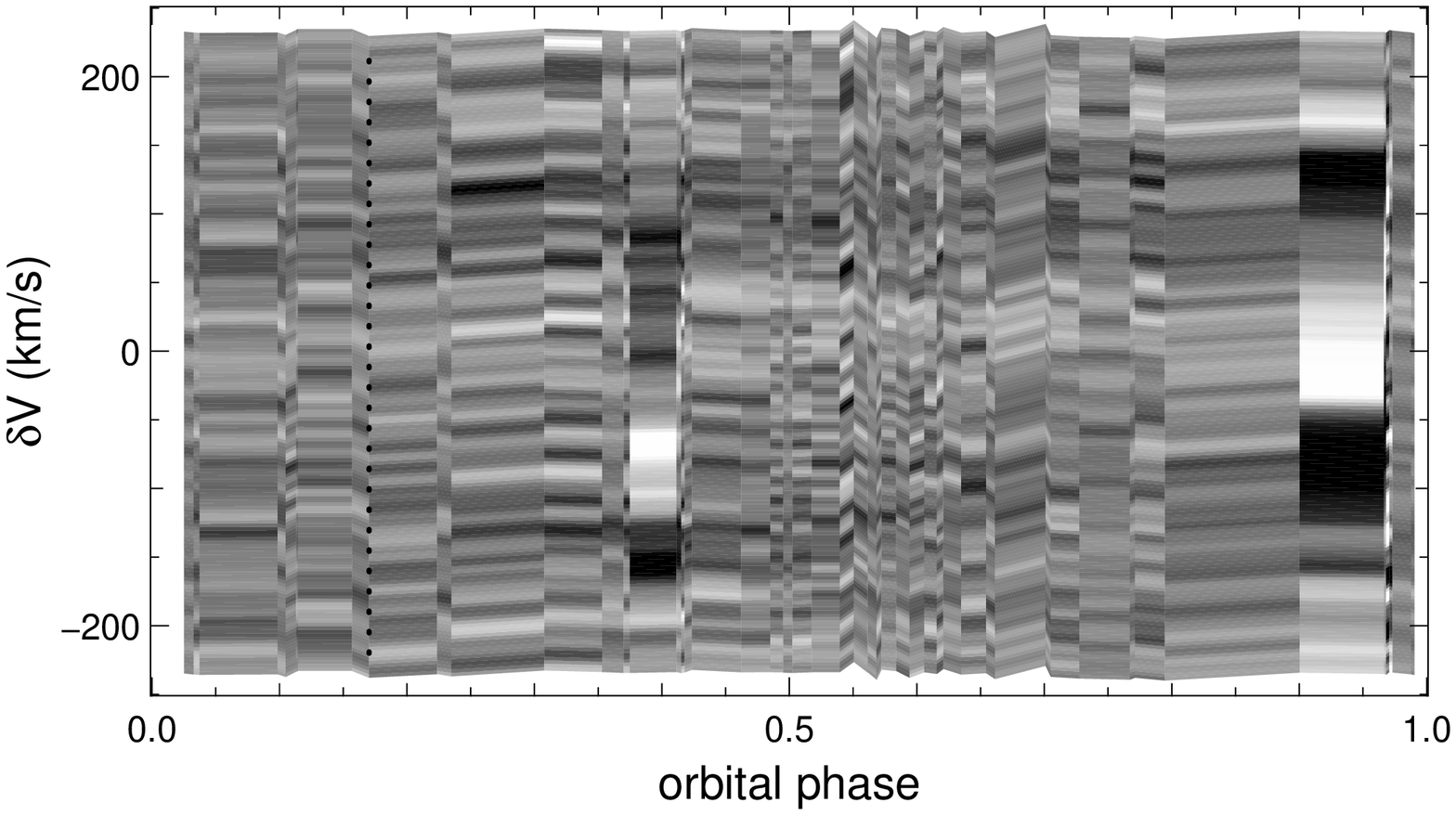}
  \caption{$\delta$~Vel Aa-Ab spectroscopic broadening functions
    (BF). \textit{Upper:} BF as a function of orbital phase;
    \textit{Lower:} residuals after the fit. See
    Fig.~\ref{fig:AaAb_BF2} for the fit at the phase corresponding to
    the dotted line ($\phi\approx0.17$). Our method to extract the
    radial velocity does not take into account the eclipses, as can be
    seen by the large residuals during the eclipses ($\phi\approx0.40$
    and $\phi \approx 0.97$, upper right panel). \label{fig:AaAb_BF1}
  }
\end{center}
\end{figure}

\begin{figure}
  \begin{center}
    \includegraphics[width=7cm]{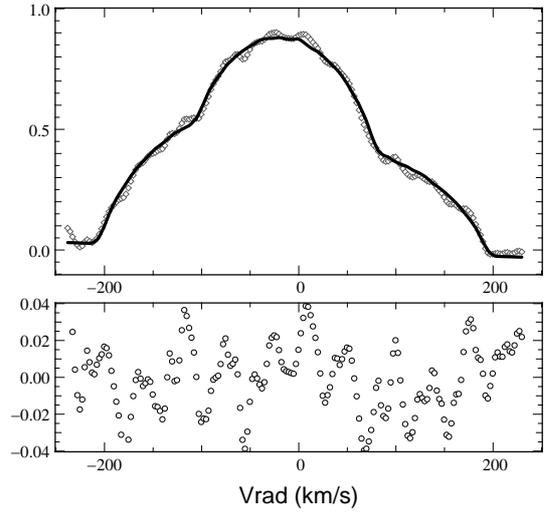}
    \caption{ Example of one fit of a broadening function (for the
      phase represented as a dotted line in Fig.~\ref{fig:AaAb_BF1},
      $\Phi\approx0.17$). The analytical fit uses the sum of two
      functions described in eq.~\ref{eq:bf}. \label{fig:AaAb_BF2}}
  \end{center}
\end{figure}

From the observed BF, it is possible to derive the $v\sin i$ from the
two components. After a few experimentations using the
stellar surface model we are going to present later, we found that the
following \textit{ad-hoc} function parameterizes well the BF for a
star seen from the equator:
\begin{equation}
\frac{\mathrm{BF}(v)}{\mathrm{BF}(v_0)} =
\left(1-\cos\left(\left[1-\left(\frac{v-v_0}{v\sin
    i} \right)^2\right]\times\pi/2\right)\right)^\alpha
\label{eq:bf}
\end{equation}
where $\alpha$ is the only parameter constraining the gravity
darkening and $v_0$ is the velocity offset. The function is defined
for $|v-v_0| \leq v \sin i$ only, and its value is 0 otherwise. Using
this analytical model and a global fit, we estimated $v\sin i$ for
each component and the radial velocities for each epoch (see
Fig.~\ref{fig:AaAb_BF1} and \ref{fig:AaAb_BF2} for the quality of the
fit). We found $v\sin i$ to be $143.5\pm0.2$~km/s and
$149.6\pm0.2$~km/s for Aa and Ab respectively. Incidentally, we find
$\alpha_a = 0.460\pm0.003$ and $\alpha_b=0.451\pm0.003$. The rotation
rate value is relatively independent of the actual gravity darkening
(parameterized here by $\alpha$) since it is set by the width of the
broadening function, not its shape.

For our fit of the orbit and the stellar parameters, we do not use the
center-to-limb darkening we derive here from the broadening
functions. The surface brightness distribution is constrained by the
photometric profil of the eclipses. However, we will check
\textit{a~posteriori} the agreement between our best fit model and the
limb-darkening derived in the analytical BF by modeling the BF from
our model. See Sect.~\ref{sec:aposteriori} and, more specifically,
Fig.~\ref{fig:aposterioriBF}.

\subsection{Global fit \label{modelfitting}}

\subsubsection{Self consistent model}

In order to extract the fundamental parameters (masses, radii, surface
temperatures, semi-major axis, etc.) from the observational data, we
propose a self consistent modeling centered around the use of physical
quantities: we model the system using two stars whose characteristics
are computed based on their radii, total luminosities and mass.

To illustrate the advantage of this approach, we can consider that in
order to model the eclipses, we could use an \textit{ad-hoc} model
based on fractional radii (ratio to the semi-major axis) and
brightness ratio, but this would not lead directly to the fundamental
parameters of the system such as effective temperatures or
luminosities. Our approach uses radii, masses and luminosities: we get
the fractional radii by self consistency between the semi-major axis
based on Newton's form of the Kepler's law (from the masses
and the period of the orbit) and the measured apparent semi-major axis
(constrained by interferometric separation vectors). The brightness
ratio arises from the luminosity and radii, and the photospheric
models we use to model the surface of the stars.

Our stellar surface model also includes stellar rotation. We model the
appearance of the star using a model developed contemporarily and
similar to the one used in Aufdenberg et al.~(\cite{aufdenberg06}) to
model the interferometric visibilities of the star Vega. To compute
the photometry, in particular during the eclipses, we generated
synthetic images and integrated them to derive the light curves. The
parameters we use are:

\begin{itemize}
\item total mass of the system (Aa+Ab);
\item the fractional mass of Aa to the total mass of the eclipsing
  system (Aa+Ab);
\item the physical radii of each component;
\item the absolute luminosity of each component;
\item the $v\sin i$ of each component, to parameterize the rotation;
\end{itemize}
In addition, we have the usual 7 parameters for the visual orbit:
\begin{itemize}
\item the period;
\item the date of passage at periastron;
\item the eccentricity;
\item three angles: inclination, $\omega$ and $\Omega$; 
\item the apparent semi-major axis (in milliarcseconds)
\end{itemize}

The only parameterization of the physics of the two components is
contained in the luminosity: the model we use of the stellar surface
is the Roche approximation, which only uses the mass, radius,
luminosity and rotational velocity. Once the shape of the
surface is computed, we link the local surface gravity and local
effective temperature using the von~Zeipel theory (von~Zeipel
\cite{vonzeipel24} and Aufdenberg et al.~\cite{aufdenberg06} for more
details). The luminosity is constrained using several mechanisms:
through the absolute photometry of the system, but also through the
surface brightness that sets the depth of the eclipses.

The advantages of using the apparent semi-major axis compared to the
physical quantity are simple. First of all, it is directly related to
one of our observables: the interferometric separation
vector. Secondly, we already have the semi-major axis by combining the
Kepler's third law, the period and the total mass. Combined with the
angular semi-major axis, we derive a distance independently from the
\emph{Hipparcos} value (a brief discussion is presented in
Sect.~\ref{distance}). The distance is derived internally in our model
and used to extract a model apparent magnitude in the K~band, which is
used as one of the constraints as we mentioned before.
\label{first_discussion_of_distance}

It is to be noted that we make the following assumptions:
\begin{itemize}
\item the stars have their rotation axis perpendicular to the plane of
  the orbit.
\item we use a Von~Zeipel gravity darkening
  coefficient of $\beta=0.25$ in $T_\mathrm{eff}\propto
  g_\mathrm{eff}^\beta$, where $T_\mathrm{eff}$ and $g_\mathrm{eff}$
  are the effective temperature and gravity at the surface. Even if
  recent observational constraints from Monnier et
  al. (\cite{monnier07}) suggest $\beta \approx 0.19$, we chose to use
  von-Zeipel's classical value since it does not lead to qualitatively
  nor quantitatively different results in our case.
\end{itemize}

\subsubsection{Orbital and stellar derived parameters}

\begin{figure}
  \begin{center}
    \includegraphics[height=7cm]{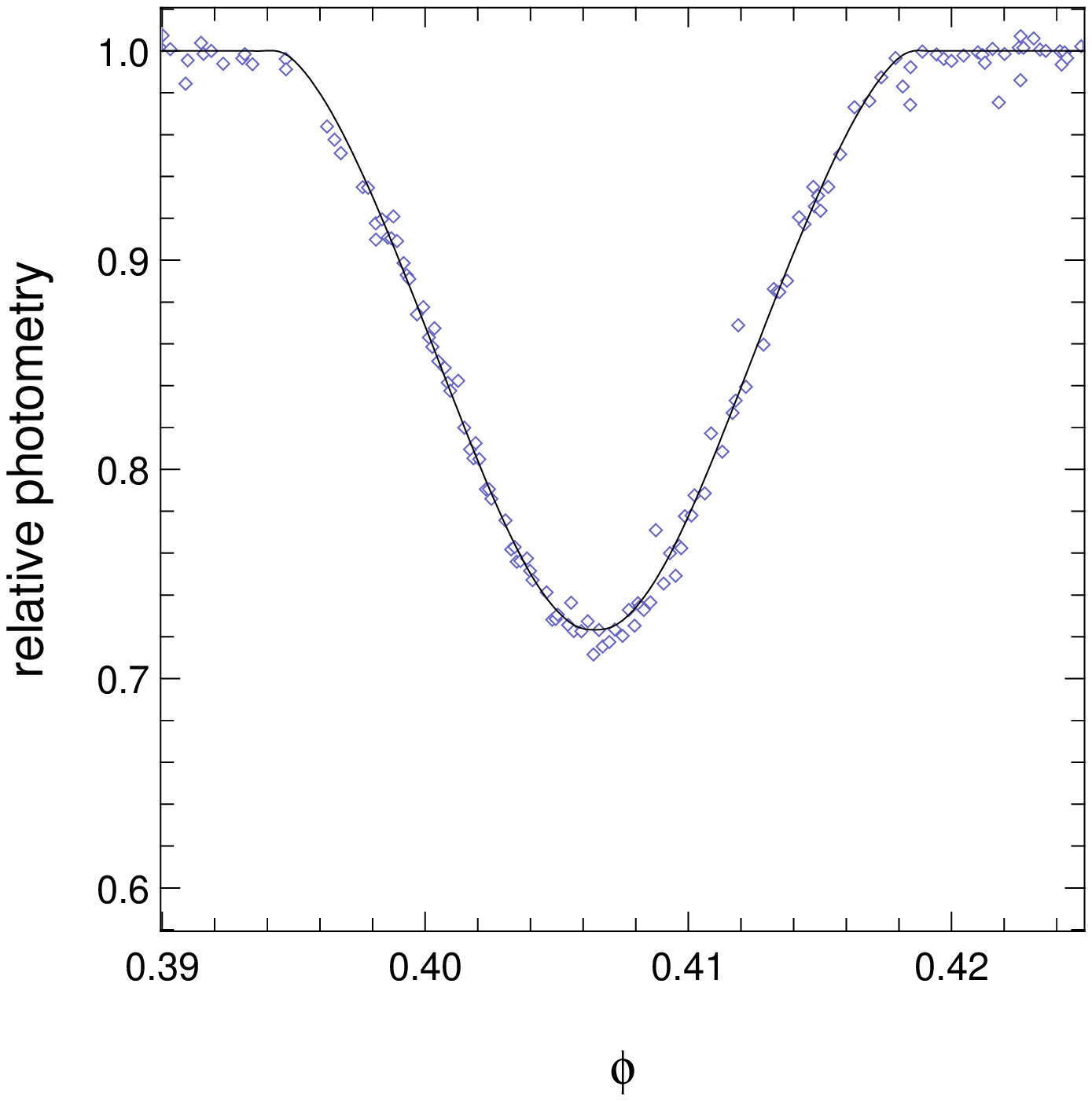}\\
    \hspace{0.5cm} \\
    \includegraphics[height=7cm]{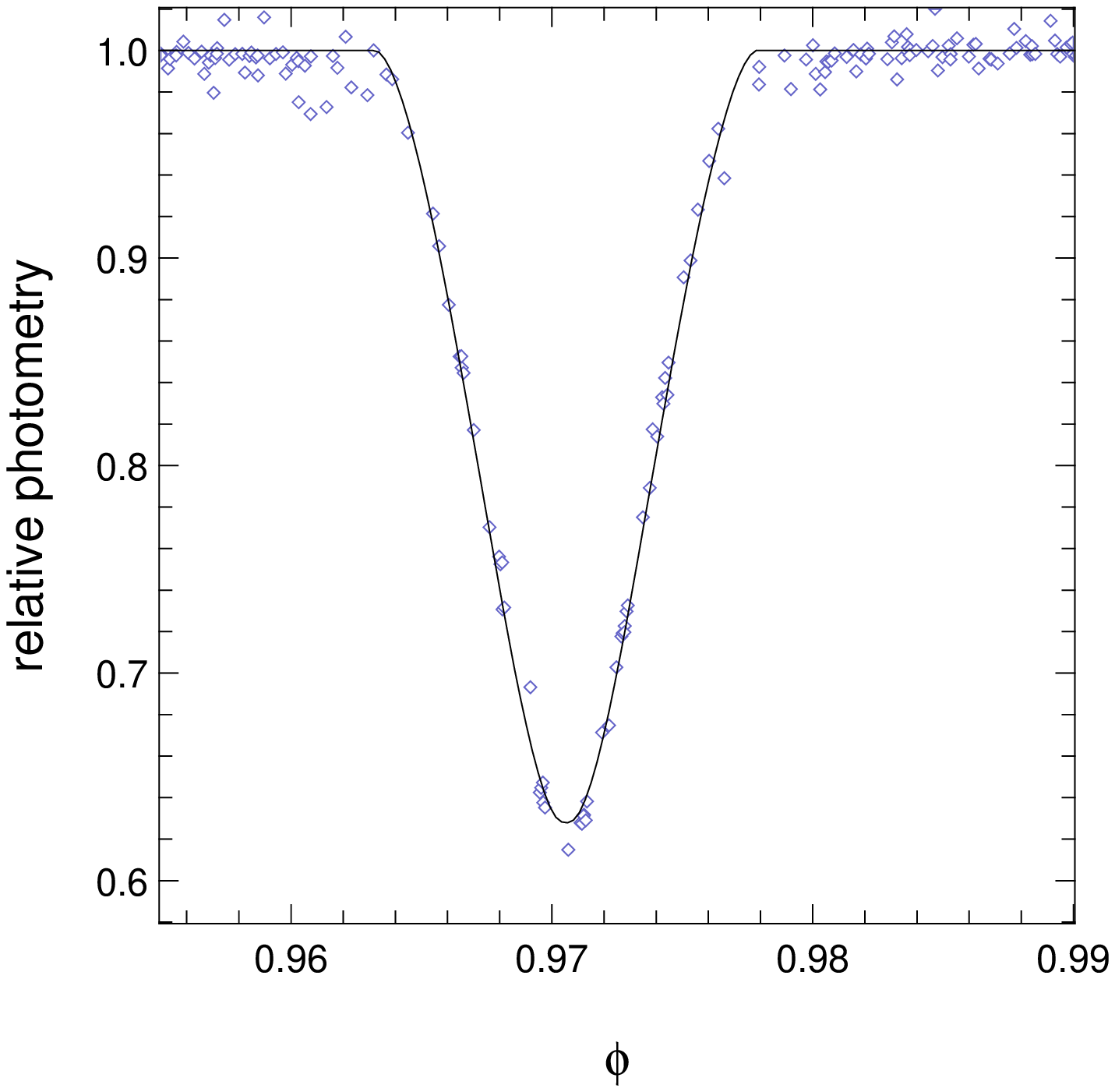}
  \end{center}
  \caption{$\delta$~Vel Aa-Ab SMEI photometric measurements (points)
    and fit of the eclipses (line) as a function of orbital phase: the
    upper panel is the the primary eclipse, the lower panel the
    secondary.\label{fig:AaAb_eclipses}}
\end{figure}

\begin{figure*}
  \begin{center}
    \includegraphics[width=12cm]{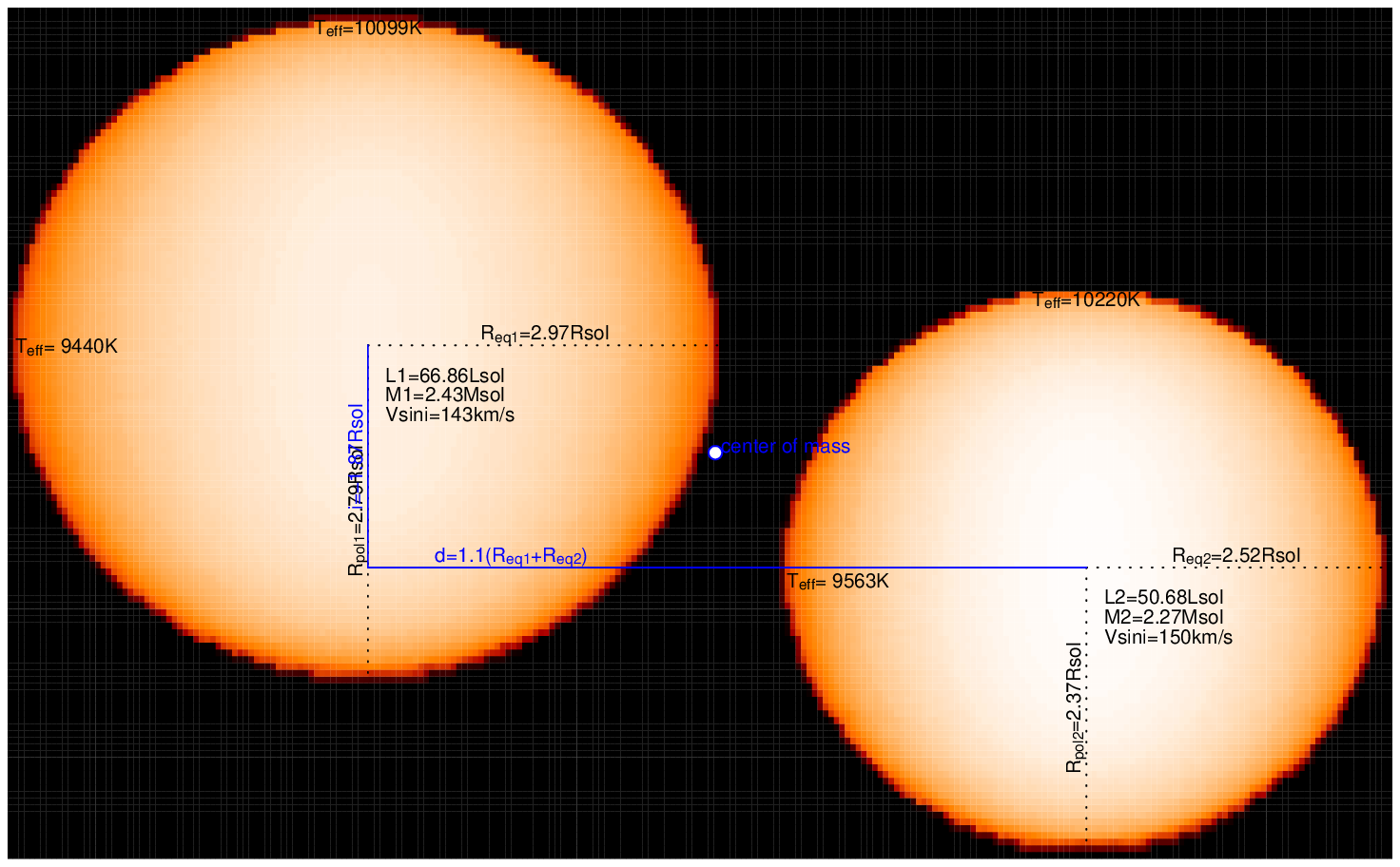}
    \includegraphics[height=7cm]{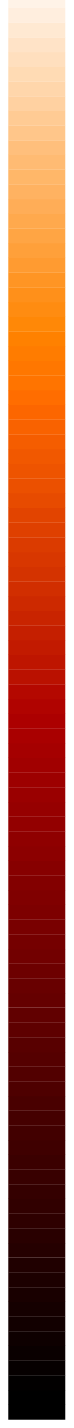}\\
    \includegraphics[width=12cm]{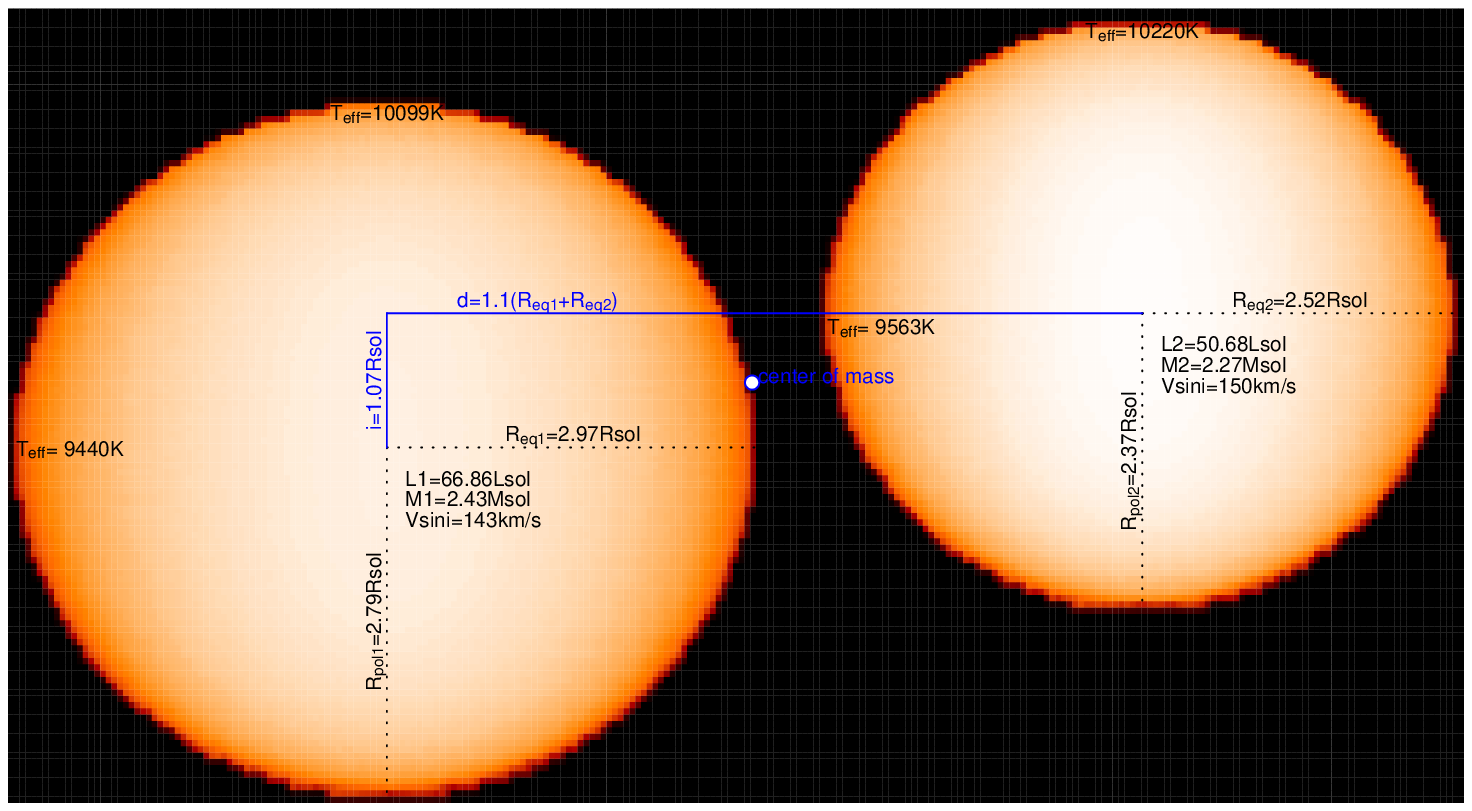}
    \includegraphics[height=6.8cm]{./figure4c.eps}\\
  \end{center}
  \caption{$\delta$~Vel Aa-Ab synthetic images in linear scale, close
    to the eclipses: \textit{the upper panel} is the configuration of
    the primary eclipse, \textit{the lower panel} is the secondary
    eclipse. \label{fig:AaAb_images}}
\end{figure*}

\begin{figure}
\begin{center}
  \includegraphics[width=8.5cm]{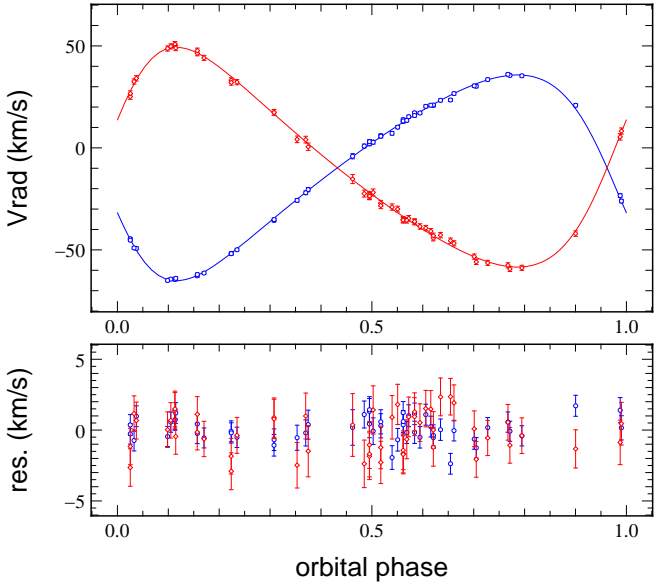}
  \caption{$\delta$~Vel Aa-Ab radial velocities as a function of
    orbital phase, with model from the fit over plotted and residuals
    plotted in the lower panel. Round points (blue) are for the
    primary and diamonds (red) for the
    secondary. \label{fig:AaAb_rad_vel}}
\end{center}
\end{figure}

\begin{figure}
  \begin{center}
    \includegraphics[width=0.4\textwidth]{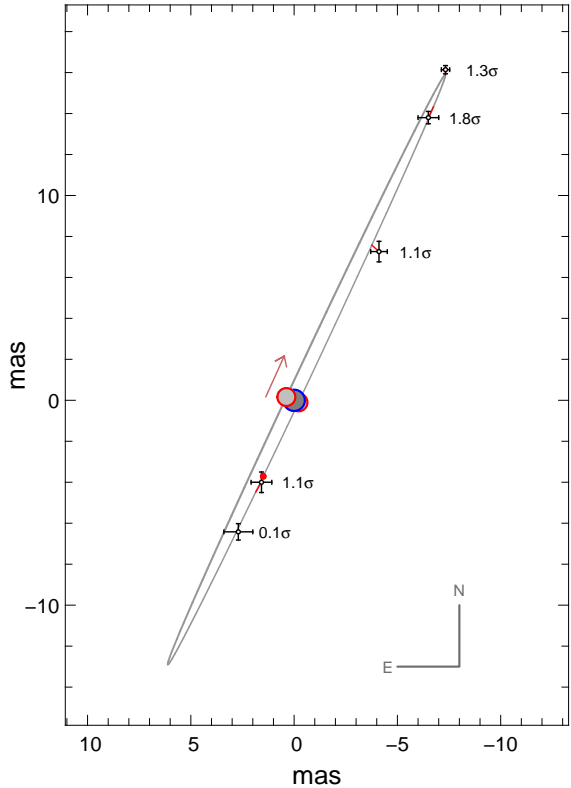}\\
  \end{center}
  \caption{$\delta$~Vel Aa-Ab visual orbit: AMBER positions (black
    dots with error bars, see Table~\ref{tab:AMBER_positions}); fitted
    orbit (gray line), Aa stellar disk (dark gray disk with blue
    surrounding), the position of Ab at periastron (red dot) and Ab
    stellar disks for the positions of the eclipses (gray disk with
    red surroundings). Residuals to the orbit (red lines, too small to
    see in most cases) and quality of the fit (in number of
    sigmas). The overall agreement corresponds to a reduced $\chi^2$
    of 1.05. \label{fig:AaAb_amber_orbit}}
\end{figure}

The result of the global fit is in excellent agreement with all the
observed data we used for the fit: the relative photometry of the
eclipses (Fig.~\ref{fig:AaAb_eclipses} and \ref{fig:AaAb_images}), the
radial velocities (Fig.~\ref{fig:AaAb_rad_vel}) and the separation
vectors (Fig.~\ref{fig:AaAb_amber_orbit}). The corresponding orbital
parameters and stellar parameters are presented in
Tables~\ref{tab:best_fit_parameters} and \ref{tab:stellar_parameters}.

\begin{table*}
  \caption{Aa-Ab orbital and stellar parameters fitted on the
    photometric, interferometric and radial velocity data, using our
    model. 
    \label{tab:best_fit_parameters}}
  \begin{center}
    \begin{tabular}{lccl}
      \hline \hline
      & Aa & Ab & constrained from\\ 
      \hline $a$ (mas) &
      \multicolumn{2}{c}{$16.51\pm0.16$ } & 
      interferometry\\ 
      Total mass ($M_\odot$) &
      \multicolumn{2}{c}{$4.69\pm0.03$} & 
      all observations
      \\
      $M_\mathrm{Aa}/(M_\mathrm{Aa}
      + M_\mathrm{Ab})$ & 
      \multicolumn{2}{c}{$0.516\pm0.001$} &
      spectroscopy\\ 
      Polar Radius ($R_\odot$) & 
      $2.79\pm0.04$ & $2.37\pm0.02$ & 
      photometry\\ 
      Luminosity ($L_\odot$)&
      $67\pm1$ & $51\pm1$ & 
      photometry \\
      $v\sin i$ (km/s) & 
      $143.5\pm0.2$ & $149.6\pm0.2$ &
      spectroscopy\\  
      Period $P$ (d) &
      \multicolumn{2}{c}{$45.1503\pm0.0002$} & 
      all observations\\      
      MJD$_0$ \textit{modulo} $P$&
      \multicolumn{2}{c}{$19.159\pm0.010$}& 
      all observations\\       
      $e$ &
      \multicolumn{2}{c}{$0.287\pm0.001$} & 
      all observations\\       
      $i$ (deg)&
      \multicolumn{2}{c}{$89.04\pm 0.03$}& 
      all observations\\ 
      $\omega$ (deg) &
      \multicolumn{2}{c}{$109.79\pm0.09$} & 
      all observations\\ 
      $\Omega$ (deg)&
      \multicolumn{2}{c}{$65.0\pm0.6$} & 
      interferometry \\ 
      $V_\gamma$ (km/s) & 
      \multicolumn{2}{c}{$-9.78\pm 0.07$} & 
      spectroscopy\\ 
      \hline
      $\chi^2$ $V_\mathrm{rad}$ Primary (err 0.75km/s) & \multicolumn{3}{c}{$1.13$}\\
      $\chi^2$ $V_\mathrm{rad}$ Secondary (err 1.3km/s) & \multicolumn{3}{c}{$1.03$}\\
      $\chi^2$ Photometry (err 0.75\%) & \multicolumn{3}{c}{$1.15$}\\
      $\chi^2$ Interferometry & \multicolumn{3}{c}{$1.05$}\\      
      \hline
    \end{tabular}
\tablefoot{The last four lines present the agreement as reduced
  $\chi^2$.}
\end{center}
\end{table*}

\begin{table}
  \caption{Derived fundamental parameters for $\delta$~Vel~Aa and
    Ab.\label{tab:stellar_parameters}. Parameters for Vega are
    displayed for comparison.}
  \begin{center}
    \begin{tabular}{llccc}
      \hline \hline
      %& \multicolumn{2}{c}{values for best fit model} \\
      &unit& $\delta$~Vel Aa & $\delta$~Vel Ab & Vega\\
      \hline
      Mass &$M_\odot$ & $2.43\pm0.02$ & 
      $2.27\pm0.02$ & $2.3\pm0.2$\\
      Luminosity &$L_\odot$ & $67\pm3$ & $51\pm2$ & $37\pm3$\\
      Polar Radius &$R_\odot$ & $2.79\pm0.04$ &
      $2.37\pm0.02$ & $2.26\pm0.07$\\
      Equ. Radius &$R_\odot$ & $2.97\pm0.02$ &
      $2.52\pm0.03$ & $2.78\pm0.02$\\
      Polar $T_\mathrm{eff.}$ &K   & 10100 & 10120 & 10150\\
      Equ.  $T_\mathrm{eff.}$ &K   & 9700 & 9560 & 7900 \\
      Avg. $T_\mathrm{eff.}$  & K  & 9450 & 9830 & 9100\\
      $\omega/\omega_\mathrm{crit.}$ & & 0.61 & 0.60 & 0.91\\
      Polar $\log(g)$ & cm s$^{-2}$& 3.90 & 4.10 & $4.10\pm0.1$ \\
      Eq. $\log(g)$ & cm s$^{-2}$& 3.78 & 3.99 & $3.65\pm0.1$ \\
      $i$ & deg & $\sim90$ & $\sim90$ & $4.7\pm0.3$ \\
      rotation rate &1/d & 0.95 & 1.17 & 1.90 \\
      %% M$_\mathrm{R}$ at $i$  & & 0.6 & 0.9 & 0.6 (1.1@90$^\mathrm{o}$)\\
      %%      apparent V mag. & & 2.6 & 2.9 & 0.03\\
      \hline
      metallicity & [M/H] & $-0.33^1$ & $-0.33^1$ & $-0.5^2$\\
      \hline
    \end{tabular}
    \tablefoot{Parameters for Vega are adapted from Aufdenberg et
      al.~(\cite{aufdenberg06}).  The references for the metallicity
      are $^1$Gray et al.~(\cite{gray06}) and $^2$Castelli \&
      Kurucz (\cite{castelli94}).}
  \end{center}
\end{table}

\subsubsection{\textit{a posteriori} verifications}
\label{sec:aposteriori}
Our model is rather simple and does not take into account one aspect:
the heating of one star by the other's radiation. This effect could be
important in our case in different conditions. The first one is if one
star was heated by the other one and developed a bright spot on its
surface. We can discard this possibility because of the absence of
photometric variations outside the eclipses. The second case where the
heating by the other star can be a problem is if this contribution is
enough to actually modify the temperature structure of the stellar
photosphere. We can check \textit{a posteriori} that the radiation
received from the other star is of the order of 3\% of the radiation
emitted (assuming similar surface brightness and a 2.5~$R_\odot$ star
seen at a distance of 89~$R_\odot$). Approximating the surface as a
blackbody, it corresponds to an increase of temperature of less than
1\% ($1.03^{1/4}$). We can thus assume that our hypothesis that
locally the photosphere can be approximated by a ATLAS model is
correct.

\begin{figure}
  \begin{center}
    \includegraphics[width=8cm]{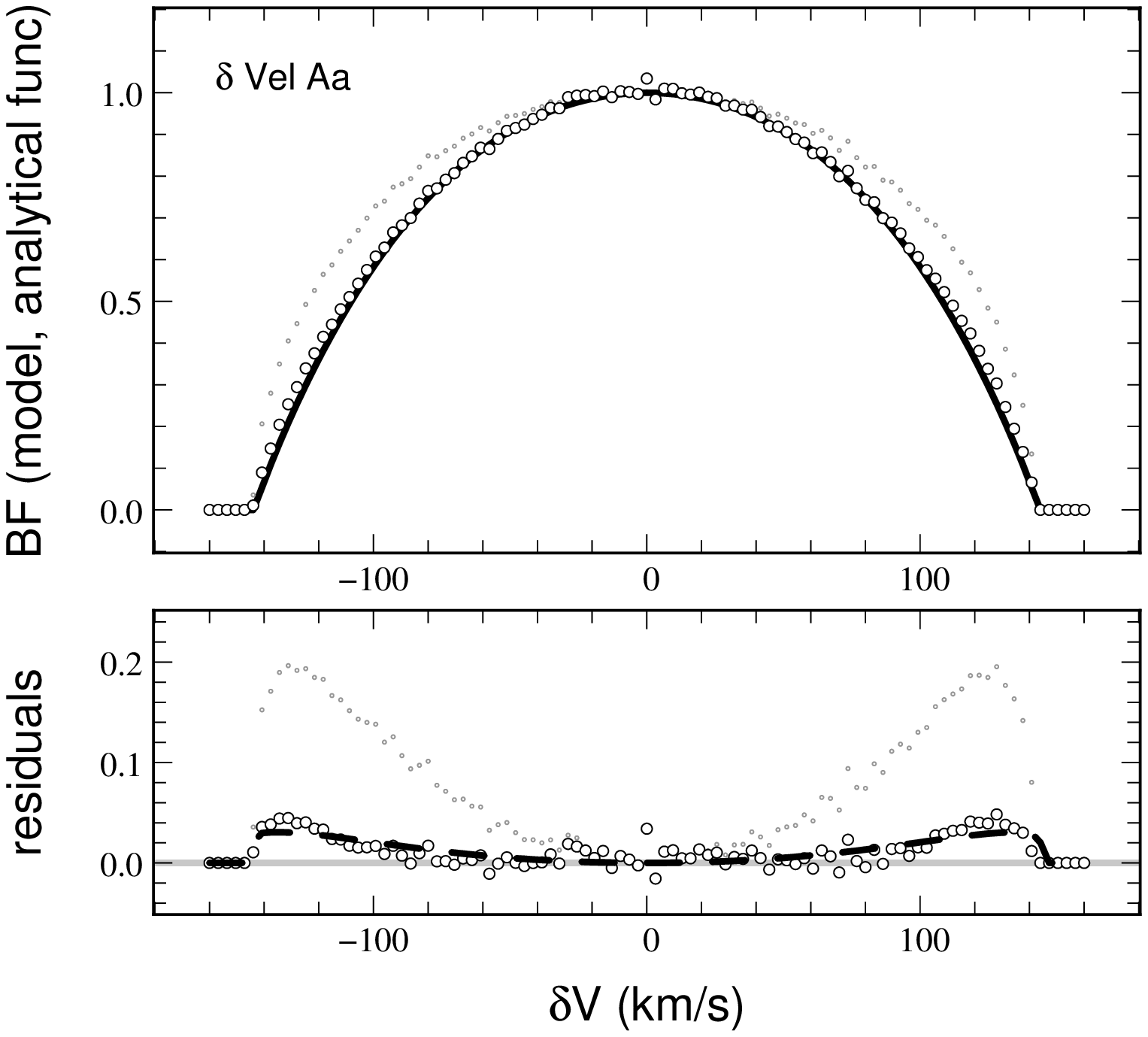}\\
    \hspace{0.2cm}
    \includegraphics[width=8cm]{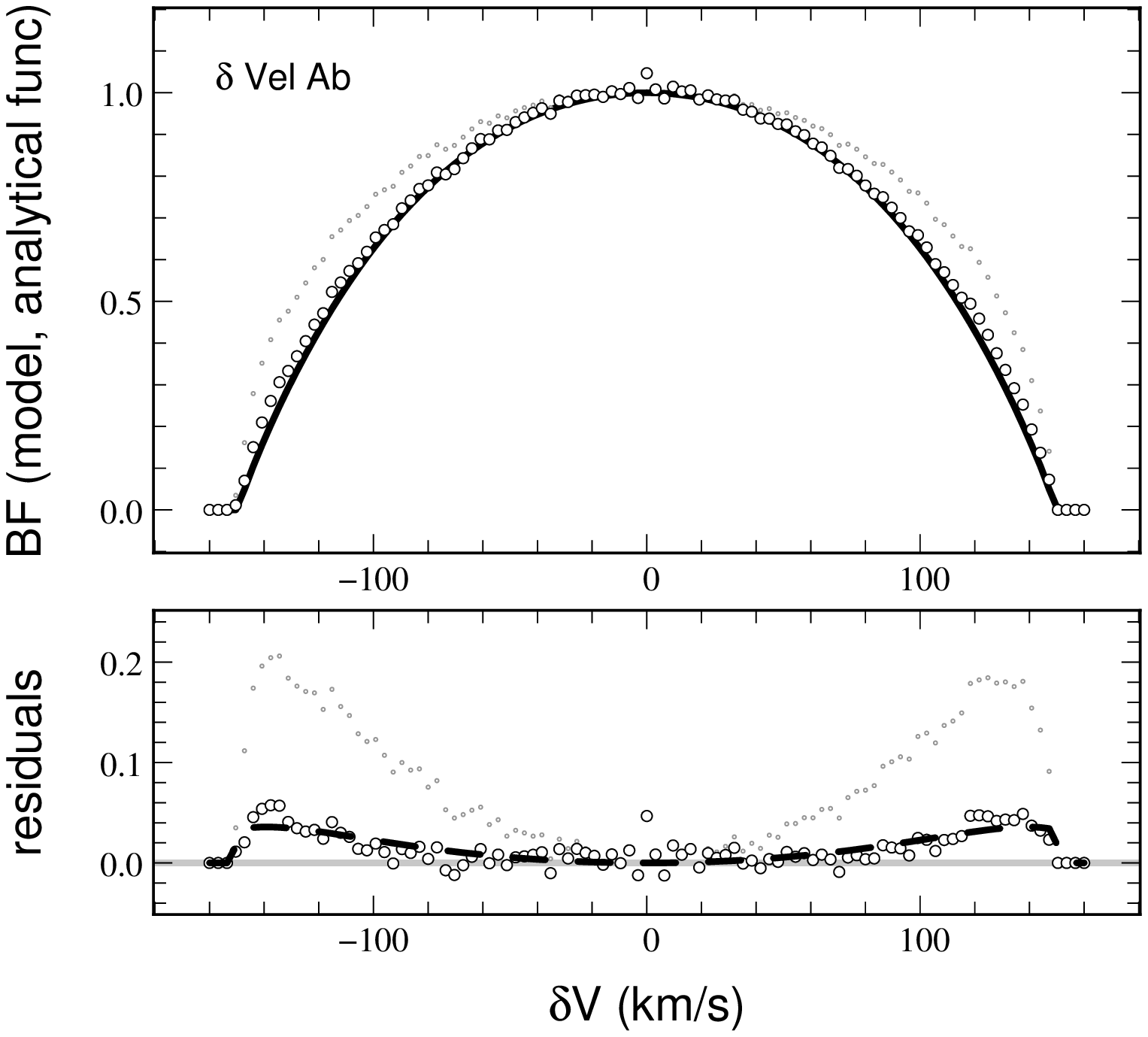}\\
  \end{center}
  \caption{$\delta$~Vel Aa-Ab expected broadening functions computed
    for our models. For each star (Aa and Ab), the open dots are for
    the model (Table~\ref{tab:stellar_parameters}) and the thick line
    is for the analytical function resulting from the fit to the
    spectroscopic data (Fig.~\ref{fig:AaAb_BF2}). The small gray dots
    are for the model as well, but ignoring the gravity darkening. In
    the smallest panels are plotted the residuals ``model'' - ''fit on
    spectroscopic data'', with the dashed line being the best fit on
    the model using the same analytical function, showing the slight
    underestimation of the gravity darkening by our model.}
  \label{fig:aposterioriBF}
\end{figure}

We can also check the consistency of our model beyond the fit of the
data we presented. In particular, we can compare the predicted
broadening functions based on our model and the broadening functions
we observe since we did not implement the direct fit of broadening
function to our model. Doing so (Fig.~\ref{fig:aposterioriBF}), the
comparison is very satisfactory, even though the wings of the data
(represented by the analytical function fit to the data in thick gray
line) seems deeper than for the model. In other words, the gravity
darkening of the model is slightly underestimated, but only by a small
fraction, considering that a model without gravity darkening (small
gray dots on Fig.~\ref{fig:aposterioriBF}) produces a very strong
disagreement.

Investigating the possible causes of this problem, we realized that we
can reproduce the more pronounced darkening of the equator compared to
the pole by tilting the star to 10 degrees from the plane of the sky:
this makes the pole more in line of sight of the observer and hence
increases the contrast between the pole and the equator.

Forcing the inclination of the spin of the stars to be 80 degrees
instead of 90 degrees does not change dramatically the fundamental
parameters of the star estimated from our fit. One of the
reasons is that it changes the sin(i) by only 1.5\%, the
actual rotational velocity is mostly unaffected. The fit converges as
well as in the case of aligned spins, with fundamental parameters
within error bars of the one estimated in the case we presented in the
main part of this work.

In conclusion of the analysis of the broadening function prediction of
our model compared to the observed one, we find a confirmation of the
consistency of our model with the data. This comparison may also
indicate that our model underestimate slightly the gravity darkening,
or, alternatively, that the stars have their rotational axis tilted on
the order of 10 degrees toward the observer, which does not impact
qualitatively nor quantitatively the fit of the data that led to the
estimation of the fundamental parameters presented earlier.

Another test is to compare the predicted wide band photometry with the
observed ones. In the literature, there is a handful of data
available, for the combined AB or A and B separately. Doing so, we see
(Table~\ref{tab:photom}) that the bluest magnitudes are not reproduced
well. Assuming a $B-V$ excess of $E(B-V)=0.055$, the difference is
nicely explained using an ISM extinction law as presented in Kervella
et al. (\cite{kervella04}).

\begin{table*}
  \caption{Computed and observed broad band photometry.}
  \label{tab:photom}
  \begin{center}
    \begin{tabular}{lccccccc}
      \hline \hline
       & \multicolumn{5}{c}{model (observation)} & $\Delta$ & extinction\\
      band & Aa & Ab & Aab & B & AB & obs.-mod. & $E(B-V)=0.055$ \\
      \hline
      B & 2.39 & 2.71 & 1.78 & 6.10 & 1.76 (\textit{2.00}$^1$)&  0.24 & 0.23\\
      V & 2.41 & 2.73 & 1.81 & 5.59 (\textit{5.54}$^2$) &  1.78 (\textit{1.96}$^1$) & 0.18 & 0.18\\
      %R & 2.47 & 2.79 & 1.86 & 5.22 & 1.82 (\textit{1.90}$^1$) & 0.18 & 0.11 \\
      %I & 2.49 & 2.82 & 1.89 & 5.01 & 1.83 (\textit{1.86}$^1$) & 0.03 & 0.10\\
      J & 2.43 & 2.76 & 1.83 & 4.72 & 1.76 (\textit{1.77$^3$}) & 0.02 & 0.04\\
      H & 2.45 & 2.78 & 1.85 & 4.44 & 1.75 (\textit{1.76$^3$}) & 0.01 & 0.03\\
      K & 2.46 & 2.78 & \textbf{1.85} (\textit{1.86$^2$}) & 4.42 (\textit{4.40}$^2$)& 1.76 (\textit{1.72$^3$}) &  -0.04 & 0.02\\
      \hline
    \end{tabular}
\tablefoot{The magnitudes predicted by the best fit model are listed
  together with the observed magnitudes (brackets). In bold is the K
  magnitude of Aab which was used to constrain our fit. The B column
  comes from the fit presented in Sect.~\ref{ABorbit}. The last two
  columns are the observed differences and the expected extinction for
  $E(B-V)=0.055$, which best fits the observed colors. Observed
  magnitudes are from Morel \& Magnenat~(\cite{morel78}) $^1$, Paper~I
  $^2$ and Skrutskie et al.~(\cite{skrutskie06}) $^3$ }
\end{center}
\end{table*}

\subsection{Distance\label{distance}}

Our parameterization allows us to derive the distance as the ratio
between of the semi-major axis of the eclipsing component (from
the total mass, the period and Kepler's third law) and its apparent
semi-major axis (from the interferometric observations). Such a
distance estimate is particularly interesting as it is purely
geometrical, and independent of the \emph{Hipparcos} measurement. From
our model, we obtain a parallax of $\pi = 39.8 \pm 0.4$\,mas for
$\delta$\,Vel.

Comparing this value to the $\pi_{\rm Hip} = 40.5 \pm 0.4$\,mas
revised \emph{Hipparcos} parallax\footnote{The revised
  \emph{Hipparcos} parallax is consistent within $1\,\sigma$ with the
  original \emph{Hipparcos} reduction ($40.9 \pm 0.4$\,mas;
  ESA~\cite{esa97}), and with the ground-based parallax of this star
  ($49.8 \pm 9.4$\,mas; Van Altena et al.~\cite{vanaltena95}).}
obtained by van Leeuwen~(\cite{vanleeuwen07}) shows a good agreement
of the two measurements, within $1.2\,\sigma$. This confirmation of
the true accuracy of these independent measurements, at the 1\% level,
shows that the \emph{Hipparcos} measurement was not disturbed by the
binary nature of $\delta$\,Vel~A. This somewhat surprising result is
due to the similar brightness ratio $L_\mathrm{Aa}/L_\mathrm{Ab}
\approx 1.3$ and mass ratio $M_\mathrm{Aa}/M_\mathrm{Ab} \approx 1.1$
of the $\delta$\,Vel~A pair. This results in a very small apparent
displacement of the center of light of the Aab system during the
orbit, with respect to the center of gravity of the two stars. Using
our model of the eclipsing system, we computed the expected
photocenter displacement during an full orbit. We find that the
peak-to-peak photocenter displacement is of the order of one
milliarcsecond, which is much smaller than the apparent astrometric
shift due to the parallax. The binarity of the system therefore did
not bias significantly the \emph{Hipparcos} parallax measurement,
neither did the low brightness of the B component. The observations of
the photocenter displacement through high-precision differential
astrometry with the VLT/NACO instrument will be the subject of a
future article.

% ############################# 2 ############################
\section{The orbit and parameters of $\delta$\,Vel~B\label{ABorbit}}
% ############################################################

\subsection{Astrometric data}

\begin{figure*}
  \begin{center}  
  \includegraphics[width=0.38\textwidth]{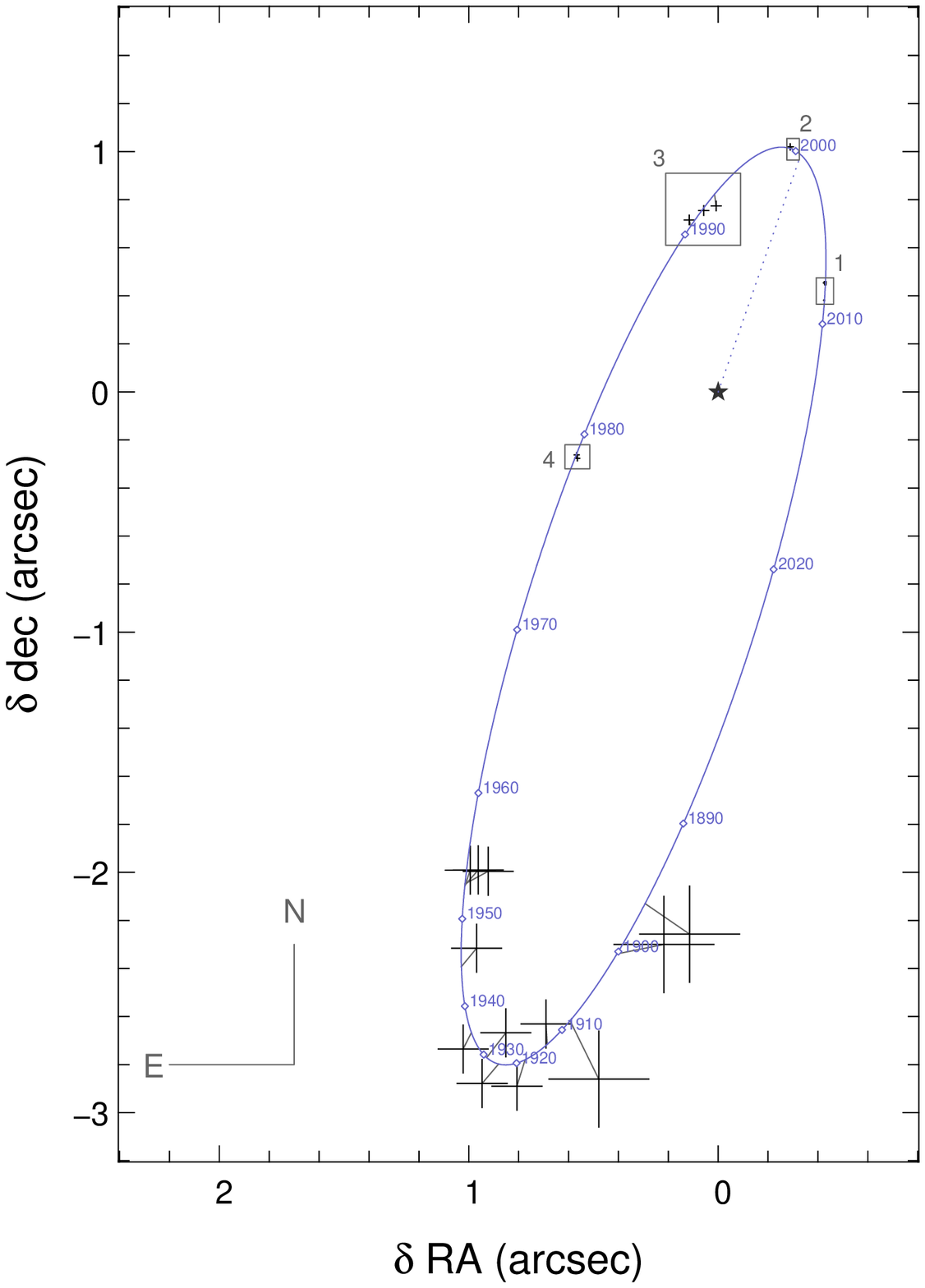}
  \hspace{0.01\textwidth}
  \includegraphics[width=0.54\textwidth]{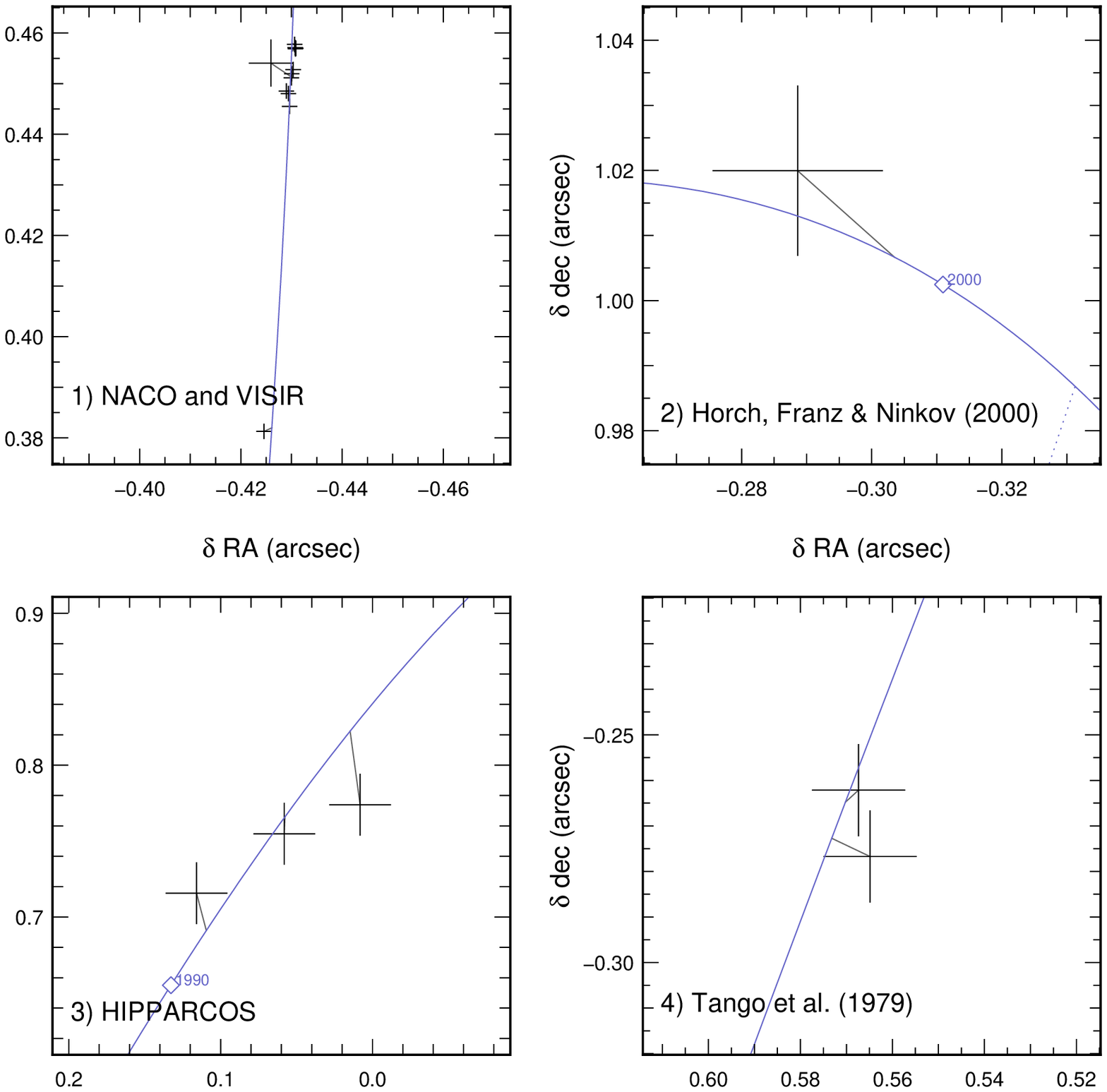} 
  \caption{\textit{left panel:} Astrometric measurements of the
    relative position of $\delta$\,Vel~B (crosses) with respect to
    $\delta$\,Vel~A ($\star$ symbol), with our best fit orbital solution (solid
    ellipse, see Table~\ref{tab:AB_orbit} for the orbital
    parameters). Thin lines connect each measurement to the
    corresponding point on the adjusted orbit. The dotted line
    corresponds to periastron passage, and the diamonds corresponds to
    the position of $\delta$\,Vel~B on Jan.\,$1^\mathrm{st}$ every ten
    years between 1890 and 2020. \textit{Right pannels (labeled 1 to
      4):} corresponds to zooming boxes on the larger view. NACO and
    VISIR data (panel 1) are tabulated in
    Table~\ref{tab:diffastrom-table}; data in panel 2 comes from
    Horch, Franz \& Ninkov (\cite{Horch00}); data in panel 4 are from
    Tango et al. (\cite{tango79}). Zooming boxes' positions are
    reported on the main (left) panel.}
  \label{fig:AB_orbit}
\end{center}
\end{figure*}

%\texttt{Note: data in an online table, I guess}

The binarity of $\delta$\,Vel was discovered by S.~I.~Bailey in 1894
from Arequipa, Peru (and independently by
Innes~\cite{innes1895}). Over more than one century, the separation
between $\delta$\,Vel~A and B has been decreasing at a rate which
nicely matches the progression of the angular resolution of the
successive generations of imaging instruments (visual observations,
photography, electronic devices). This progression allowed a
relatively regular tracing of the visual orbit of the pair, down to
the sub-arcsecond separations that occur around the periastron
passage. With the advent of speckle interferometry (Tango et
al.~\cite{tango79}) and the \emph{Hipparcos} satellite
(ESA~\cite{esa97}) the accuracy of the measured relative positions
improved significantly. In Paper~I, we present in details the new data
we obtained with the \emph{Very Large Telescope}, using both the
K-band adaptive optic system VLT/NACO (Rousset et
al.~\cite{rousset03}; Lenzen et al.~\cite{lenzen98}) and the N-band
camera VLT/VISIR (Lagage et al.~\cite{lagage04}). Thanks to the large
aperture of the telescopes and the diffraction-limited angular
resolution, these observations provide us with new high-precision
astrometry of the A-B pair. The resulting separations of
$\delta$\,Vel~B relatively to A are presented in
Table~\ref{tab:diffastrom-table}. For the conversion of the separation
measured in pixels to angular separations, we adopted the pixel scale
of $13.26 \pm 0.03$\,mas/pixel (Masciadri et al.~\cite{masciadri03})
for NACO and $75 \pm 1$\,mas/pixel for VISIR. The assumed NACO plate
scale is in good agreement with the calibration by Neuh\"auser et
al.~(\cite{neuhauser08}), who demonstrated that this figure is stable
over a period of at least 3 years. The VISIR plate scale uncertainty
is set arbitrarily to $\approx 1\%$, although it is probably better in
reality. The angular separation was only $\approx 0.6\arcsec$ for the
epoch of our observations. 

In addition to these new astrometric measurements, we also take
advantage of the historical astrometric positions assembled by Argyle
et al.~(\cite{argyle02}) in his Table~5, that includes 17 epochs
between 1895 and 1999. It is to be noted that these authors used the
two speckle interferometry epochs from Tango et al.~(\cite{tango79})
with a different definition for the projection angle, leading to an
apparent inconsistency with the other measurements. Transforming the
Tango et al. projection angle $PA$ using $PA\rightarrow(180-PA)$,
these two data points become much more consistent with the other
epochs and observing techniques.

\begin{table}
\caption{Differential astrometry of $\delta$\,Vel from NACO and VISIR
  images.
\label{tab:diffastrom-table}}
\begin{center}
\begin{tabular}{cccc}
\hline \hline
UT date & MJD-54\,000 & $[\alpha_B - \alpha_A]$ & $[\delta_B - \delta_A]$ \\
& & mas & mas \\
\hline
\noalign{\smallskip}
2008-04-01$^N$ & 557.0224 & $ -430.6 \pm 1.0 $ & $ 457.8 \pm 1.0 $ \\
2008-04-04$^N$ & 560.9976 & $ -430.7 \pm 1.0 $ & $ 457.1 \pm 1.0 $ \\
2008-04-06$^N$ & 562.0121 & $ -430.8 \pm 1.0 $ & $ 457.1 \pm 1.0 $ \\
2008-04-07$^N$ & 563.0048 & $ -430.7 \pm 1.0 $ & $ 457.0 \pm 1.0 $ \\
2008-04-20$^N$ & 576.9715 & $ -430.3 \pm 1.0 $ & $ 452.9 \pm 1.0 $ \\
2008-04-23$^N$ & 579.0231 & $ -430.0 \pm 1.0 $ & $ 452.0 \pm 1.0 $ \\
2008-04-24$^V$ & 580.0503 & $ -425.9 \pm 4.3 $ & $ 454.0 \pm 4.5 $ \\
2008-04-24$^N$ & 580.9917 & $ -429.8 \pm 1.0 $ & $ 451.1 \pm 1.0 $ \\
2008-05-05$^N$ & 591.9748 & $ -428.9 \pm 1.0 $ & $ 448.7 \pm 1.0 $ \\
2008-05-07$^N$ & 593.9732 & $ -429.2 \pm 1.0 $ & $ 448.0 \pm 1.0 $ \\
2008-05-18$^N$ & 604.0442 & $ -429.6 \pm 1.0 $ & $ 445.4 \pm 1.0 $ \\
2009-01-07$^N$ & 838.1347 & $ -424.6 \pm 1.0 $ & $ 381.3 \pm 0.9 $ \\
\hline
\end{tabular}
\end{center}
\tablefoot{ measurements from NACO have symbol $^N$, whereas
  measurements from VISIR have $^V$. The angles are all expressed in
  milliarcseconds.}
\end{table}

\subsection{Orbital elements}

We adjusted the orbital parameters of the $\delta$\,Vel~A-B pair to
the whole sample of astrometric data, and the result is presented
graphically in Fig.~\ref{fig:AB_orbit}. The corresponding orbital
elements are listed in Table~\ref{tab:AB_orbit}. It should be noted
that thanks to a semi-major axis twice more precise, and a period ten
time more precise, the total mass value derived from based Kepler's
third law is significantly improved, which becomes limited by our
parallax estimation of $\pi = 39.8 \pm 0.4$.

\begin{equation}
M(\mathrm{Aab+B}) = 6.15 \pm 0.15_\mathrm{orbit} \pm
0.17_\mathrm{parallax}\,M_\odot
\label{eq:AB_mass}
\end{equation}

\begin{table}
\caption{Orbital parameters of the visual pair A-B.}
\label{tab:AB_orbit}
\begin{center}
  \begin{tabular}{lcc}
    \hline \hline
      parameter & this work & Argyle et al.~(\cite{argyle02})\\
      \hline
      $a$ ($''$)          & $1.996\pm0.012$ & $1.990\pm 0.020$\\
      $e$              & $0.475\pm0.003$ &  $0.470\pm0.020$ \\
      %Period (d)      & $52314\pm 450$ &  $51865.50\pm 4748$ \\
      Period (yr)      & $143.2\pm1.2$   & $142\pm13$\\
      $\mathrm{MJD}_0$ & $51774\pm 430$  & $51836.80\pm584$ \\
      $i$ (deg)        & $105.1\pm  0.2$ &  $105.20\pm  2.20$ \\ 
      $\Omega$ (deg)   & $286.6\pm 0.36$ & $287.00\pm  1.30$ \\ 
      $\omega$ (deg)   & $187.4\pm  0.6$ & $188.00\pm 14.00$ \\
      \hline
      $M(\mathrm{Aab+B})$ ($M_\odot$) & 
      $6.15\pm0.23$ & $5.88\pm 1.17$\\
      \hline
    \end{tabular}
\end{center}
\tablefoot{The total mass (last line) is computed using Kepler's third
  law, assuming the revised parallax $\pi = 39.8 \pm 0.4$\,mas. }
\end{table}

\subsection{Physical properties of $\delta$\,Vel~B}

We used a spectral energy distribution (hereafter SED) fit to the
photometric data (Table~\ref{tab:B_SED}) corrected for reddening
assuming $E(B-V)=0.055$. We used a carefully interpolated grid of ATLAS
models (e.g. Kurucz~\cite{kurucz05}) in order to derive the angular
diameter and effective temperature of $\delta$\,Vel~B. We find a
photospheric limb darkened angular diameter of $\theta_\mathrm{LD}(B)
= 0.530 \pm 0.011$\,mas and an effective temperature of
$T_\mathrm{eff}(B) = 6600 \pm 100$\,K. Based on our distance estimate,
we can derive the physical radius to be $R(B) = 1.43 \pm
0.03$\,$R_\odot$ and thus a luminosity of $L(B) = 3.5 \pm
0.2\,L_\odot$. Assuming the star is on the main sequence we
can infer, based on the mass-luminosity relation by Torres, Andersen
\& Gim\'enez (\cite{torres10}), that $\delta$\,Vel~B has a spectral
type F7.5V and a mass of:
\begin{equation}
 M_\mathrm{photometric}(B) = 1.35 \pm 0.1\,M_\odot
\label{eq:mb_phot}
\end{equation}
These parameters estimated using an independent method are comparable
to the values we obtained in Paper~I.

\begin{table}
  \caption{Spectral Energy Density (SED) fit of $\delta$\,Vel~B, using
    our photometric measurements in the K and N bands (in two narrow
    band filters of VISIR, PAH1 and PAH2).}
  \label{tab:B_SED}
    \begin{center}
      \begin{tabular}{ccccc}
        \hline \hline
        filter & meas. & redd. & SED & modeled SED\\
        & & & $\mathrm{W} / \mathrm{m}^2 /
      \mu\mathrm{m}$ & $\mathrm{W} / \mathrm{m}^2 /
      \mu\mathrm{m}$ \\
        \hline
        V (Mag)  & 5.54 & 0.18 &  2.159$\times 10^{-10}$  & $2.143\ 10^{-10}$ \\
        K (Mag) & 4.40 & 0.02 & 6.951$\times 10^{-12}$ & $7.135\ 10^{-12}$ \\
        PAH1 (flux) & 0.94Jy & negl.& 3.816$\times 10^{-14}$ & $3.816\  10^{-14}$ \\
        PAH2 (flux) & 0.58Jy  & negl.& 1.369$\times 10^{-14}$ & $1.310\  10^{-14}$ \\
        \hline
      \end{tabular}
    \tablefoot{The third column is the reddening correction
      corresponding to $E(B-V)=0.055$. The model (last column) is for
      an angular diameter of 0.53~mas and effective temperature of
      6600K.}
    \end{center}
\end{table}

We can compare this mass estimate with the value we can compute from
the mass of A ($4.69\pm0.03M_\odot$ from
table~\ref{tab:best_fit_parameters}) and A+B ($6.15\pm0.23M_\odot$
from table~\ref{tab:AB_orbit}. This leads to:
\begin{equation}
  M_\mathrm{dynamic}(B) = M(A+B)-M(A)=1.46\pm0.23\,M_\odot
\end{equation}
which is consistent with the photometric estimate
(Eq.~\ref{eq:mb_phot})

% ############################# 5 ############################
% \section{Discussion}
% ############################################################

%\subsection{Fundamental parameters and evolutionary stage}
%
%On Fig.~\ref{fig:AaAb_HR}, we placed the two stars Aa and Ab in the
%H-R diagram as well as evolutionary tracks (Marigo et al. 2008) for
%$\delta$~Vel: $[M/H]=-0.33$, that is Z=0.009 (Gray et al. 2006). We
%found that a best agreement is reached if we use a grid corresponding
%to Z=0.008 (Solar correpsonds to Z=0.019). Assuming we can compare the
%stars to non-rotating evolution model, we can estimate the age of
%$\delta$~Vel Aa and Ab to be of the order of $450\pm50$~Myr, which is
%very amazingly close to the estimated age of Vega: 390 to 570 Myr
%(Peterson et al. 2006). 
%
%\begin{figure}
%\begin{center}
%  \includegraphics[width=8cm]{./FIGURES/AaAb_HRdiagram_Z0.008.eps}
%  \caption{$\delta$~Vel Aa-Ab H-R diagram, with evolutionnary tracks
%    and isochrones for metal deficient compared to the Sun (Z=0.012),
%    from Marigo et al. (2008). The numbers indicate the age of the
%    model, in millions of years from the main sequence. For each star,
%    the horizontal line is the range of temperature (from equatorial
%    to polar), the dot is for the average temperature. We remind here
%    that we estimated $M_\mathrm{Aa}=2.43\,M_\odot$ and
%    $M_\mathrm{Ab}=2.27\,M_\odot$, in perfect agreement with the
%    isochrones. \label{fig:AaAb_HR}}
%\end{center}
%\end{figure}

% ############################# 6 ############################
\section{Conclusion}

We presented a self consistent model of the triple stellar system
$\delta$\,Vel. Our model reproduces photometric, spectroscopic and
interferometric data of the eclipsing pair Aa-Ab. We determined the
orbital (Table~\ref{tab:best_fit_parameters}) and fundamental stellar
parameters (Table~\ref{tab:stellar_parameters}) of the three
components of the system. The physical properties of the eclipsing
components are surprisingly similar to the A0V benchmark star
Vega. Thanks to the resolution of the system using the AMBER
instrument, we also independently determine the distance to the system
($\pi = 39.8 \pm 0.4$\,mas, or $25.1 \pm 0.25$\,pc), as well as the
interstellar reddening value towards this nearby system, with
$E(B-V)=0.055$. The combination of two fast rotating A stars of
slightly different masses and a late F star, all coeval and with
accurately measured fundamental parameters, will likely make of
$\delta$\,Vel a cornerstone for the study of early-type main sequence
stars.
% ############################################################

%__________________________________Acknowledgements
\begin{acknowledgements}
T.P. acknowledges support from the European Union in the FP6 MC ToK
project MTKD-CT-2006-042514. This work has partially been supported by
VEGA project 2/0094/11. This work also received the support of PHASE,
the high angular resolution partnership between ONERA, Observatoire de
Paris, CNRS and University Denis Diderot Paris 7. This research has
made use of the \texttt{AMBER data reduction package} of the
Jean-Marie Mariotti Center\footnote{Available at
  http://www.jmmc.fr/amberdrs}. This research took advantage of the
SIMBAD and VIZIER databases at the CDS, Strasbourg (France), and
NASA's Astrophysics Data System Bibliographic Services.
\end{acknowledgements}

%__________________________________Bibliography
{}


\begin{thebibliography}{}
\bibitem[1985]{aumann85} Aumann, H. H. 1985, PASP, 97, 885
\bibitem[2006]{absil06} Absil, O., Di Folco, E., M\'erand, A., et
  al. 2006, A\&A, 452, 237
\bibitem[2002]{argyle02} Argyle, R. W., Alzner, A., \& Horch,
  E. P. 2002, A\&A, 384, 171
\bibitem[2006]{aufdenberg06} Aufdenberg, J. P., M\'erand, A., Coud\'e
  du Foresto, V., et al. 2006, ApJ, 645, 664, Erratum 2006, ApJ, 651,
  617
%\bibitem[1989]{cardelli89} Cardelli J., Clayton G. \& Mathis J, 1989,
%  ApJ 345, 245
\bibitem[1994]{castelli94} Castelli, F., \& Kurucz, R. L. 1994, A\&A,
  281, 817
%\bibitem[1999]{cox99} Cox A., et al. 1999, Astrophysical Quantities,
%  ISBN 978-0387987460
\bibitem[2009]{chelli09} Chelli, A., Utrera, O. H. \& Duvert, G. 2009,
  A\&A, 502, 705
\bibitem[1997]{esa97} ESA 1997, The Hipparcos and Tycho Catalogues,
  ESA SP-1200
\bibitem[2008]{gaspar08} G\'asp\'ar, A., Su, K. Y. L., Rieke, G. H.,
  et al. 2008, ApJ, 672, 974
\bibitem[2006]{gray06} Gray, R. O., Corbally, C. J., Garrison, R. F.,
  et al. 2006, AJ, 132, 161
\bibitem[1999]{grenier99} Grenier, S., Burnage, R., Farragiana,
  R. 1999, A\&AS. 135, 503
\bibitem[1994]{gulliver94} Gulliver, A. F., Hill, G., Adelman, S. J.
  1994, ApJ, 429, 81
\bibitem[2003]{hempel03} Hempel, M., \& Schmitt, J. H. M. M. 2003,
  A\&A, 408, 971
\bibitem[2000]{Horch00} Horch, E; Franz, O. G. \&
  Ninkov, Z. 2000, AJ 120, 2638 
\bibitem[1895]{innes1895} Innes, R. T. A. 1895, MNRAS, 55, 312
\bibitem[2007]{kellerer07} Kellerer, A., Petr-Gotzens, M., Kervella,
  P., \& Coud\'e du Foresto, P. 2007, A\&A, 469, 633
\bibitem[2004]{kervella04} Kervella, P., Bersier, D., Mourard, D. et
  al. 2004, A\&A, 428, 587
\bibitem[2009]{kervella09} Kervella, P., Th\'evenin, F., \&
  Petr-Gotzens M. G. 2009, A\&A, 493, 107 (Paper I)
\bibitem[2005]{kurucz05} Kurucz, R. L. 2005, MmSAI Suppl., 8, 14
\bibitem[2004]{lagage04} Lagage, P.O. et al. 2004, The ESO Messenger 117, 12
\bibitem[1998]{lenzen98} Lenzen, R., Hofmann, R., Bizenberger, P., \&
  Tusche, A. 1998, SPIE 3354, 606
\bibitem[2003]{masciadri03} Masciadri, E., Brandner, W., Bouy, H. et
  al. 2003, A\&A, 411, 157
\bibitem[2005]{merand05} M\'erand, A., Bord\'e, P., \& Coud\'e du
  Foresto, V. 2005, A\&A, 433, 1155
\bibitem[2010]{moerchen10} Moerchen, M. M., Telesco, C. M., \&
  Packham, C. 2010, ApJ, 723, 1418
\bibitem[2007]{monnier07} Monnier J.-D., Zhao M., Pedretti E., et
  al. 2007, Sci, 317, 342
\bibitem[1978]{morel78} Morel M., \& Magnenat P. 1978, A\&A Suppl.,
  34, 477
\bibitem[2008]{neuhauser08} Neuh\"auser, R., Mugrauer, M., Seifahrt,
  A., Schmidt, T. O. B., \& Vogt, N. 2008, A\&A, 484, 281
\bibitem[2000]{otero00} Otero, S. A., et al. 2000, IBVS, 4999
\bibitem[2007]{petrov07} Petrov, R. G., Millour, F., Chelli, A., et
  al. 2007, A\&A, 464, 1
\bibitem[2011]{pribulla11} Pribulla, T., M\'erand, A., Kervella, P.,
  et al. 2011, A\&A, 528, A21 (Paper II)
\bibitem[2003]{rousset03} Rousset, G., Lacombe, F., Puget, F., et
  al. 2003, Proc. SPIE 4839, 140
%\bibitem[2008]{seifahrt08} Seifahrt, A., R\"oll, T., \& Neuh\"auser,
%R. 2008, Proc. 2007 ESO Instrument Calibration Workshop, Garching,
%Germany, 271
\bibitem[2006]{skrutskie06} Skrutskie, R. M., Cutri, R., Stiening,
  M. D., et al. 2006, AJ, 131, 1163
\bibitem[2008]{spreckley08} Spreckley, S. A., \& Stevens, I. R. 2008,
  MNRAS, 388, 1239
\bibitem[2006]{su06} Su, K. Y. L., Rieke, G. H., Stapelfeldt, K. R.,
  et al 2008, ApJ, 679, L125
\bibitem[1979]{tango79} Tango, W. J., Davis, R. J., Thompson, R. J.,
  \& Hanbury Brown, R. 1979, Proc. Australian Astro. Soc., 3, 323
\bibitem[2007]{tatulli07} Tatulli, E., Millour, F., Chelli, A., et
  al. 2007, A\&A, 464, 29
\bibitem[2010]{torres10} Torres, G., Andersen, J. \and Giménez, A 2010,
   A\&ARv 18, 67
\bibitem[1995]{vanaltena95} Van Altena, W. F., Lee, J. T., \&
  Hoffleit, E. D. 1995, The General Catalogue of Trigonometric
    Stellar Parallaxes, 4$^{\rm th}$ Edition (Yale University
  Observatory)
\bibitem[2007]{vanleeuwen07} van Leeuwen, F. 2007, {Hipparcos, the New
  Reduction of the Raw Data}, Astrophysics and Space Science Library,
  (Springer, Berlin) 350
\bibitem[1924]{vonzeipel24} von~Zeipel, H. 1924, MNRAS, 84, 665
\end{thebibliography}
\end{document}